\documentclass[11pt,a4paper]{article}
\usepackage[utf8]{inputenc}
\usepackage[english]{babel}
\usepackage{amsmath}
\usepackage{amsthm}
\numberwithin{equation}{section}
\usepackage{amsfonts}
\usepackage{mathtools}
\usepackage{mathrsfs}
\usepackage{mathpazo}
\usepackage{multirow}
\usepackage{amssymb}
\usepackage{graphicx}
\usepackage{ifpdf}
\usepackage{braket}
\newcommand{\changefont}[3]{
\fontfamily{#1} \fontseries{#2} \fontshape{#3} \selectfont}
\changefont{ppl}{m}{n}

\usepackage{a4wide}

\usepackage{cite}
\usepackage[bookmarks=true,colorlinks=true,linkcolor=black,citecolor=black,urlcolor=black,bookmarksnumbered]{hyperref}

\begin{document}

\setcounter{equation}{0}
\setcounter{footnote}{0}
\setcounter{section}{0}

\thispagestyle{empty}
\vspace{5.0truecm}
\begin{center}

{\Large \bf Lax pairs for new $\mathbb{Z}_N$-symmetric coset $\sigma$-models 

and their Yang-Baxter deformations}

\vspace{1.0truecm}

{David Osten}

\vspace{1.0truecm}

{\em Institute for Theoretical and Mathematical Physics \\
Lomonosov Moscow State University \\

Lomonosovsky Avenue, Moscow, 119991, Russia}

\vspace{1.0truecm}

{{\tt ostend@itmp.msu.ru}}

\vspace{1.0truecm}
\end{center}

\begin{abstract}
Two-dimensional $\sigma$-models with $\mathbb{Z}_N$-symmetric homogeneous target spaces have been shown to be classically integrable when introducing WZ-terms in a particular way. This article continues the search for new models of this type now allowing some kinetic terms to be absent, analogously to the Green-Schwarz superstring $\sigma$-model on $\mathbb{Z}_4$-symmetric homogeneous spaces. A list of such integrable $\mathbb{Z}_N$-symmetric (super)coset $\sigma$-models for $N \leq 6$ and their Lax pairs is presented. For arbitrary $N$, a big class of integrable models is constructed that includes both the known pure spinor and Green-Schwarz superstring on $\mathbb{Z}_4$-symmetric cosets.

Integrable Yang-Baxter deformations of this class of $\mathbb{Z}_N$-symmetric (super)coset $\sigma$-models can be constructed in same way as in the known $\mathbb{Z}_2$- or $\mathbb{Z}_4$-cases. Deformations based on solutions of the modified classical Yang-Baxter equation, the so-called $\eta$-deformation, require deformation of the constants defining the Lagrangian and the corresponding Lax pair. Homogeneous Yang-Baxter deformations (i.e. those based on solutions to the classical Yang-Baxter equation) leave the equations of motion and consequently the Lax pair invariant and are expected to be classically equivalent to the undeformed model.

As an example, the relationship between $\mathbb{Z}_3$-symmetric homogeneous spaces and nearly (para-)K\"ahler geometries is revisited. Confirming existing literature it is shown that the integrable choice of WZ-term in the $\mathbb{Z}_3$-symmetric coset $\sigma$-model associated to a nearly K\"ahler background gives an imaginary contribution to the action.

\end{abstract}

\pagebreak
\tableofcontents

\section{Introduction}
Non-linear $\sigma$-models \cite{GellMann:1960np} appear ubiquitously in particle physics, condensed matter theory and gravity. In two dimension they can be classically integrable. Prototypical example are the principal chiral model or the symmetric space $\sigma$-model \cite{Eichenherr:1979ci}. Applied to string theory, supersymmetric versions exist: integrable coset $\sigma$-models with supergravity backgrounds as target space \cite{Henneaux:1984mh,Metsaev:1998it,Berkovits:1999zq,Bershadsky:1999hk,Bena:2003wd,Vallilo:2003nx,Arutyunov:2008if,Stefanski:2008ik}. These supergravity backgrounds are very unique spaces -- flat space, Lie supergroups, AdS$_n\times$S$^n$ or AdS$_4 \times \mathbb{CP}^3$. There are two directions in which one could go to use the advantages of integrability in more generic models: examples with less symmetry and examples with (NSNS-)fluxes, in particular for the application in string theory. Corresponding modifications or deformations, that one could make at the level of the action, are to introduce WZ-terms or to perform a Yang-Baxter deformation \cite{Klimcik:2002zj,Klimcik:2008eq,Delduc:2013qra,Hollowood:2014rla,Kawaguchi:2014fca,Klimcik:2015gba,Matsumoto:2015jja,Hoare:2016hwh}. This article will be partly concerned with latter. These deformations are generated so-called $R$-operators -- algebraic objects based on an underlying Lie algebra structure: solution to the classical or the modified classical Yang-Baxter equation. In former case these are typically called \textit{homogeneous Yang-Baxter deformation}, in latter \textit{$\eta$-deformation}. Homogeneous Yang-Baxter deformations are closely connected to abelian and non-abelian $T$-duality transformations. \cite{Matsumoto:2014nra,Osten:2016dvf,Hoare:2016wsk,Borsato:2016pas,Borsato:2017qsx,Fernandez-Melgarejo:2017oyu,Sakamoto:2017cpu,Hoare:2017ukq,Lust:2018jsx,Hoare:2018ngg}. $\eta$-deformations, on the other hand, are non-trivial deformations of the classical dynamics and have novel target space interpretations \cite{Delduc:2013fga,Delduc:2014kha,Hoare:2015wia,Arutyunov:2015mqj,Hoare:2016ibq,Borsato:2016ose,Orlando:2016qqu,Borsato:2018spz,Hoare:2018ngg,vanTongeren:2019dlq}. It is also possible to combine deformations based on WZ-terms and seperate Yang-Baxter deformations of the left and right action of the underlying group into multiparametric integrable deformations \cite{Klimcik:2014bta,Delduc:2014uaa,Hoare:2014oua,Delduc:2015xdm,Klimcik:2017ken,Delduc:2017fib,Delduc:2018xug,Klimcik:2019kkf,Seibold:2019dvf,Hoare:2020mpv}. Structures appearing have some similarities to a problem posed in this paper, the interplay of having WZ-terms and performing a Yang-Baxter deformation.

A different way to generalise from the setting of a symmetric space $\sigma$-model and its supersymmetric version, is to consider $\sigma$-models with $\mathbb{Z}_N$-symmetric homogeneous target spaces, generalising the $\mathbb{Z}_2$- resp. $\mathbb{Z}_4$-symmetry of the former. \cite{Young:2005jv,Ke:2008zz,Ke:2011zzb,Ke:2011zzc,Ke:2011zzd,Bykov:2014efa,Bykov2016cyclic,Bykov:2016rdv,Ke:2017wis,Bykov:2018vbg,Delduc:2019lpe,Hoare:2021dix}. Based on a $\mathbb{Z}_N$-grading of an underlying Lie algebra $\mathfrak{g} = \bigoplus_i \mathfrak{g}^{(i)}$, allows for existence of certian WZ-terms. These need to be introduced in order to make the $\sigma$-model on the $\mathbb{Z}_N$-symmetric coset space integrable. Moreover it is has been shown that, in principle, $\mathbb{Z}_{2N}$-symmetric spaces can be supergravity backgrounds, also for $N >2$ \cite{Kagan:2005wt}. The central question of this article is: What other choices of integrable $\sigma$-models on $\mathbb{Z}_N$-symmetric homogeneous spaces are there and do all of these possess an integrable Yang-Baxter deformation?

\paragraph{Key results.} The central object of interest in this article is the action
\begin{equation}
S = \frac{1}{2} \int \mathrm{d}^2\sigma \ \text{Tr}(j_+ \mathcal{P}_- j_-) \sim \int \sum_{i=1}^{N-1} \text{Tr}(s_i j^{(i)} \wedge \star j^{(N-i)} + b_i j^{(i)} \wedge j^{(N-i)} )
\end{equation}
for a group-valued field $g(\sigma)$ and a projector $\mathcal{P}_- = \sum_i (s_i - b_i) \mathcal{P}^{(i)}$ with $\mathcal{P}^{(i)}$ being the projector to the $\mathfrak{g}^{(i)}$-component. This model is written in terms of its Maurer-Cartan-current $j = g^{-1} \mathrm{d}g$ and characterised by constants $s_i$ and $b_i$ in front of kinetic resp. WZ-terms corresponding to the $\mathfrak{g}^{(i)}$-component of the $\mathbb{Z}_N$-decomposition. Generalising known results in the literature \cite{Young:2005jv,Ke:2008zz,Hoare:2021dix} one choice for these constants, such that the model is classically integrable, is
\begin{align}
s_i &= 0 = s_{N-i} = 0 , \qquad b_{i} = - b_{N-i} = - \frac{2i}{N} \qquad \text{for} \quad i=1,...,K, \\
s_i &= 1 \qquad b_i = 1 - \frac{2i}{N} \qquad \text{for} \quad i=K+1,...,N-K-1.
\end{align}
for a $K \in \{1,...,M\}$ where $N = 2M+1$ resp $2M+2$. Such different integrable choices of the $s_i \in \{0,1 \}$ were to be expected, in analogy to the existence of the pure spinor and Green-Schwarz superstring in the case of $\mathbb{Z}_4$-(super)cosets. Other choices are discussed in section \ref{chap:New}. This setting can be Yang-Baxter deformed, as in certain cases already demonstrated in \cite{Ke:2017wis,Hoare:2021dix},
\begin{equation}
S = \frac{1}{2} \int \mathrm{d}^2 \sigma \ \text{Tr} \left( j_+ \mathcal{P}_- \frac{1}{1 - \eta R_g \circ \mathcal{P}_-} j_- \right) .
\end{equation}
This model is shown to be classically integrable for certain choices of the constants $s_i$, $b_i$. Here $R_g = \text{Ad}_g^{-1} R \text{Ad}_g$. $R$ is a linear operator on $\mathfrak{g}$ and solution of the (modified) classical Yang-Baxter equation. Such Yang-Baxter deformations of $\mathbb{Z}_N$-cosets are constructed in section \ref{chap:YangBaxter}. \pagebreak

\section{Review: the pure spinor-type $\mathbb{Z}_N$-coset $\sigma$-model}
\label{chap:Review}
This section reviews the original construction \cite{Young:2005jv} and some applications of the integrable $\mathbb{Z}_N$-cosets $\sigma$-model before generalising it in several directions in the following sections.

\subsection{Action and Lax Connection}
We consider a $\sigma$-model with target space $M = G/H$, where $G$ and $H$ are Lie (super)groups with Lie (super)algebras $\mathfrak{g}$ resp. $\mathfrak{h} \equiv \mathfrak{g}^{(0)}$ that are subject to a $\mathbb{Z}_N$-grading
\begin{align}
\mathfrak{g}  = \mathfrak{h} \oplus \mathfrak{m} &= \mathfrak{g}^{(0)} \oplus \mathfrak{g}^{(1)} \oplus ... \oplus \mathfrak{g}^{(N-1)} \\
\text{with} \qquad [\mathfrak{g}^{(i)},\mathfrak{g}^{(j)}] &\subset \mathfrak{g}^{(i+j \ \text{mod} N)}.
\end{align}
with $\mathfrak{m} = \bigoplus_i \mathfrak{g}^{(i)}$. In other words we have a Lie algebra automorphism $\sigma: \ \mathfrak{g}\rightarrow \mathfrak{g}$ with $\sigma^N = 1$ and the $\mathfrak{g}^{(i)}$ being the eigenspaces to that automorphism. 

For a non-Abelian algebra $\mathfrak{g}$ and with $N >2$
\begin{equation}
[\mathfrak{h} , \mathfrak{h}] \subset \mathfrak{h}, \quad [\mathfrak{h},\mathfrak{m}] \subset \mathfrak{m} \quad \text{and} \quad [\mathfrak{m},\mathfrak{m}] \not\subset \mathfrak{h}.
\end{equation}
The last relation makes the difference to a symmetric homogeneous space (a $\mathbb{Z}_2$-coset), explaining names like \emph{non-symmetric, semi-symmetric} or \emph{$N$-symmetric homogeneous space} for these in the literature. In this paper these spaces are referred to as \textit{$\mathbb{Z}_N$-cosets} for brevity. 

These spaces might seem very abstract, but in general they have a close connection to almost Hermitian manifolds \cite{Wolf1968HomogeneousSD,GrayNearlyKaehler} and have been studied in the mathematical literature. A canonical example is $M = (K \times K \times ... \times K) / K_{diag} = K^N/K_{diag}$ for a Lie group $K$ and with the corresponding $\mathbb{Z}_N$-automorphism being the permutation of the $N$ copies of $K$. An explicit geometric interpretation for $\mathbb{Z}_3$-symmetric homogeneous spaces in terms of nearly (para-)K\"ahler geometry is given in section \ref{chap:Applications}. 

Furthermore, the existence of a non-degenerate Ad-invariant bilinear form on $\mathfrak{g}$, respecting its $\mathbb{Z}_N$-grading, is required in order to construct a $G$-invariant action:
\begin{equation}
\text{Tr}\left( m n \right) = \sum_{i = 0}^{N-1}\text{Tr}\left( m^{(i)} n^{(N-i)} \right) \quad \text{with} \quad m^{(i)},n^{(i)} \in \mathfrak{g}^{(i)}.
\end{equation}
Here it is denoted by 'Tr', but it can be any bilinear form with these properties. All calculations in this paper hold as well for Lie superalgebras with a $\mathbb{Z}_{2N}$-grading, as in \cite{Kagan:2005wt}, where it would be the supertrace.

\paragraph{Lagrangian.} A non-linear $\sigma$-model on a two-dimensional worldsheet $\Sigma$ with light-cone coordinates $\sigma_\pm = (\tau \pm \sigma)/2$ on such a $\mathbb{Z}_N$-coset  $M = G/H$ may be defined by the Lagrangian
\begin{equation}
\mathcal{L} = \frac{1}{2} \text{Tr} \left(j_+ \mathcal{P}_- (j_-) \right) = \frac{1}{2} \text{Tr} \left(\mathcal{P}_+(j_+) j_- \right) , \label{eq:Lagrangian}
\end{equation}
with the Maurer-Cartan currents $j = g^{-1} \mathrm{d} g$ of a group valued field $g: \ \Sigma \rightarrow G$, and the operators
\begin{align}
\mathcal{P}_\pm = \sum_{i=1}^{N-1} (s_i \pm b_i) \mathcal{P}^{(i)}
\end{align}
with constants $s_i = s_{N-i}$ and $b_i = - b_{N-i}$, for $i = 1,...,N-1$ and where $\mathcal{P}^{(i)}$ is the projector to $\mathfrak{g}^{(i)}$. The constants $s_i$ and $b_i$ characterise the model. I.e. the Lagrangian can be written as
\begin{equation}
\mathcal{L} \sim \sum_i \text{Tr}\left( s_i j^{(i)} \wedge \star j^{(N-i)} + b_i j^{(i)} \wedge j^{(N-i)} \right),
\end{equation}
where $s_i$ and $b_i$ are the factors in front of kinetic resp. WZ-terms. The existence of these WZ-terms in that form relies on the existence of the $\mathbb{Z}_N$-automorphism. The Lagrangian respects the 'target space' symmetries of the $\mathbb{Z}_N$-cosets in the following way.
\begin{itemize}
\item global (right) $G$-symmetry $g \mapsto g_0\cdot g$, for (constant) $g_0 \in G$, of the Maurer-Cartan forms $j$ itself
\item local (left) $H$-symmetry $g \mapsto g \cdot h$, where $h^{-1} \mathrm{d} h \in \mathfrak{h}$,
\begin{equation}
j^{(i)} \mapsto \left(h^{-1} j h + h^{-1} \mathrm{d}h \right)^{(i)} = (h^{-1} j h)^{(i)} = h^{-1} j^{(i)} h
\end{equation}
The last equality followed from $[\mathfrak{h},\mathfrak{g}^{(i)}] \subset \mathfrak{g}^{(i)}$. With the Ad-invariance of the bilinear form this implies the invariance of the Lagrangian.
\end{itemize}
In addition there is a mapping $b_i \leftrightarrow - b_i$ and $\mathfrak{g}^{(i)} \leftrightarrow \mathfrak{g}^{(N-i)}$  for all $i$, that will leave the action invariant.

\paragraph{The equations of motion} are given by
\begin{equation}
\mathcal{E} = \partial_+ \left(\mathcal{P}_- j_- \right) + \partial_- \left(\mathcal{P}_+ j_+ \right) + [j_+ , \mathcal{P}_- j_-] + [j_-, \mathcal{P}_+ j_+] = 0. \label{eq:ZNEOM}
\end{equation}
These and the Maurer-Cartan condition for $j = g^{-1} \mathrm{d}g$,
\begin{equation}
\mathcal{M} = \partial_+ j_- - \partial_- j_+ + [j_+,j_-] = 0,
\end{equation} 
become, when decomposed in the grading eigenspaces ($i = 1,...,N-1$), 
\begin{align}
\mathcal{M}^{(0)} &= \partial_+ j_-^{(0)} - \partial_- j_+^{(0)} + [j_+^{(0)} , j_-^{(0)}] + \sum_{j=1}^{N-1} [j_+^{(j)} , j_-^{(N-j)}] = 0  \nonumber \\
\mathcal{M}^{(i)} &= \partial_+ j_-^{(i)} - \partial_- j_+^{(i)} + [j_+^{(0)} , j_-^{(i)}] + [j_+^{(i)} , j_-^{(0)}] + \sum_{j\neq 0,i} [j_+^{(j)} , j_-^{(i-j)}] = 0 \label{eq:ZNEOMDecomp}\\
(\mathcal{E}^{(i)} + b_i \mathcal{M}^{(i)}) &= s_i (\partial_+ j_-^{(i)} + \partial_- j_+^{(i)} + [j_+^{(0)} , j_-^{(i)}] - [j_+^{(i)} , j_-^{(0)}]) + \sum_{j\neq 0,i} \mathcal{D}_{ij} [j_+^{(j)} , j_-^{(i-j)}] = 0 \nonumber
\end{align}
with constants
\begin{align}
\mathcal{D}_{ij} &= \left\lbrace \begin{array}{cr}
f_{N+j-i} - f_{j} - b_i & i>j \\
f_{j-i} - f_{j} - b_i  & i<j \end{array}\right. . \label{eq:ZNEOMDecompConst}
\end{align}
Here the combination $f_i = s_i - b_i$ was introduced such that
\begin{align*}
\mathcal{P}_+ = \sum_{i=1}^{N-1} f_i \mathcal{P}^{(i)}, \qquad \mathcal{P}_- = \sum_{i=1}^{N-1} f_{N-i} \mathcal{P}^{(i)}.
\end{align*}
\paragraph{The 'pure spinor'-type model and its Lax integrability.}
The original work \cite{Young:2005jv} considers the case, in which $s_i\neq 0$ for all $i = 1, ... ,N-1$. In that case the equations of motion can be represented as a Lax pair. Putting an ansatz for the Lax connection like
\begin{equation}
L_\pm = \sum_{i=1}^{N-1} l^{(\pm)}_i j^{(i)}_\pm,
\end{equation}
with constants $l^{(\pm)}_i$, into the flatness condition $\partial_+ L_- - \partial_- L_+ + [L_+ , L_-] = 0$, using the Maurer-Cartan identity and the equations of motion \eqref{eq:ZNEOMDecomp}, one obtains a set of equations for the constants $l_i^{(\pm)}$. This system becomes underdetermined\footnote{There is also a second solution with opposite sign in that condition. The correspond model and its Lax connection can be obtained by the previously discussed mapping $b_i \rightarrow -b_i, \ \mathfrak{g}^{(i)} \rightarrow \mathfrak{g}^{(N-i)}$.}, when for all $i,j = 1,...,N-1$ with $i\neq j$:
\begin{align}
\mathcal{D}_{ij} &= \left\lbrace \begin{array}{cr}
s_i & i>j \\
-s_i  & i<j \end{array}\right.  \quad \Rightarrow \quad 0=\left\lbrace \begin{array}{cr}
f_{N+j-i} - f_{j} - f_{N-i} & i>j \\
f_{j-i} - f_{j} + f_i  & i<j \end{array}\right. .
\end{align}
This way one finds a choice for the constants $f_i = s_i - b_i$ rendering the corresponding model classically integrable
\begin{align}
f_i = s \frac{2i}{N} \quad \Rightarrow \quad s_i = s, \ b_i = s \left(1 - \frac{2i}{N} \right) \ \forall i, 
\end{align}
in the case that all $s_i \neq 0$. Consequently, all kinetic terms are present in the action \eqref{eq:Lagrangian}. $s$ is a free constant in front of the action and not interesting for the classical considerations in this paper, and hence we set $s=1$. One parameterisation of the corresponding Lax connection with spectral parameter $\lambda$ is given by
\begin{equation}
L_+(\lambda) = j^{(0)}_+ + \sum_{i=1}^{N-1}\lambda^i j^{(i)}_+, \qquad L_-(\lambda) = j^{(0)}_- + \sum_{i=1}^{N-1}\frac{1}{\lambda^{N-i}} j^{(i)}_-.
\end{equation}
As observed in \cite{Young:2005jv,Kagan:2005wt}, for $N=4$ this Lax pair and the Lagrangian \eqref{eq:Lagrangian} correspond to the ones of the pure spinor-superstring as obtained, for example, for AdS$_5 \times$S$^5$ \cite{Vallilo:2003nx}. For that reason, we will call the case of $s_i = 1, \ i=1,...,N-1$ the \textit{pure spinor-type} $\mathbb{Z}_N$-coset $\sigma$-model in the following.

Another case of an integrable $\sigma$-model on a $\mathbb{Z}_N$-coset, known in the literature, is the $\mathbb{Z}_4$-supercoset $\sigma$-model of \textit{Green-Schwarz type} \cite{Henneaux:1984mh,Metsaev:1998it,Arutyunov:2008if,Stefanski:2008ik}, meaning that the fermionic components $j^{(1)}$ and $j^{(3)}$ do not possess kinetic terms. Hence, that case corresponds to $s_1 = s_3 = 0$. One generalisation of that to $\mathbb{Z}_N$-(super)cosets with $N > 4$ has been given in \cite{Ke:2008zz} for $N = 4M$, with for $s_{2M} = 1$ and all others $s_i = 0$. Generalising both of these classes, new models of \textit{hybrid type}\footnote{hybrid between the \textit{pure spinor}- and \textit{Green-Schwarz}-type, meaning some fermionic or bosonic kinetic terms are set to zero.} will be introduced in section \ref{chap:New}, in which $s_i = 0$ only for some $i$. The nomenclature, \textit{pure spinor-}, \textit{Green-Schwarz}- or hybrid type will be used, although one might not work with a supercoset, for example for odd $N$.

\subsection{Applications}
\label{chap:Applications}
In this section, besides collecting some basic information on $\mathbb{Z}_N$-cosets as supergravity backgrounds, the connection of $\mathbb{Z}_3$-cosets to homogeneous nearly K\"ahler- resp. para-K\"ahler structures is reviewed. In the mathematical literature this connection has been pointed out long ago and studied in detail \cite{Wolf1968HomogeneousSD,GrayNearlyKaehler,Butruille2005}. In the physics literature on $\sigma$-models this connection and the correspondic $\sigma$-models have appeared in \cite{Bykov:2014efa,Bykov2016cyclic,Bykov:2016rdv,Bykov:2018vbg,Delduc:2019lpe}. In particular, it is pointed out here in the following that the $\sigma$-models based on nearly K\"ahler geometry might be interesting, but that tuning the parameter such that it the $\sigma$-model is integrable requires an complex action. On para-complex target spaces one might be able to avoid these issues \cite{Delduc:2019lpe}.

\subsubsection{On the connection of $\mathbb{Z}_3$-cosets and nearly (para-)K\"ahler geometries}
Let $M = G/H$ where $\mathfrak{g} = \mathfrak{h} \oplus \mathfrak{m}$ is a real semisimple Lie algebra. The eigenvalues of a $\mathbb{Z}_3$-automorphism $\sigma: \ \mathfrak{g} \rightarrow \mathfrak{g}$, are complex: $1,\mu,\mu^2$ with $\mu^3$ = 1. The problem, that the eigenspaces are in general subspaces of $\mathfrak{g}^\mathbb{C}$, can be dealt with by observing that the subspaces $\mathfrak{h}$ and $\mathfrak{m}$ are real: $\mathfrak{m} = (\mathfrak{m}_\mu \oplus \mathfrak{m}_{\mu^2}) \cap \mathfrak{g}$.
The connection to an \textbf{almost complex structure} $\mathbb{J}$ is that restricted to $\mathfrak{m}$ one can write the $\mathbb{Z}_3$-automorphism $\sigma^3 = 1$ as
\begin{equation}
\sigma\vert_{\mathfrak{m}} = -\frac{1}{2} \mathbb{1} + \frac{\sqrt{3}}{2} \mathbb{J} \quad \text{with} \quad \mathbb{J}^2 = -1. \label{eq:Z3ComplexStructureRel}
\end{equation}
$\mathfrak{g}^{(1)}$ and $\mathfrak{g}^{(2)}$ are the holomorphic/antiholomorphic eigenspaces to $\mathbb{J}$, which can be defined on the (real) algebra $\mathfrak{m}$ (i.e. without referring to $\mathfrak{g}^\mathbb{C}$). Under complex conjugation $\overline{\mathfrak{g}^{(1)}} = \mathfrak{g}^{(2)}$. With a metric $\langle \cdot , \cdot \rangle$ on $M$, that can be derived from the Ad-invariant bilinear form on $\mathfrak{g}$ as for any homogeneous space, one can form the compatible tripe $(\langle \cdot , \cdot \rangle, \mathbb{J} , \omega$ with the nearly K\"ahler form $\omega$, with $\mathrm{d} \omega \neq 0$ that gives rise non-trivial WZ-term in the corresponding $\sigma$-model below. $\mathbb{J}$ is subject to the following conditions:
\begin{equation}
\mathbb{J}[m_1,m_2]\vert_{\mathfrak{m}} = - [\mathbb{J}m_1 , m_2]\vert_{\mathfrak{m}} = - [m_1 , \mathbb{J}m_2]\vert_{\mathfrak{m}}, \qquad [\mathbb{J}m_1 , m_2]\vert_{\mathfrak{h}} = - [m_1 , \mathbb{J}m_2]\vert_{\mathfrak{h}} \label{eq:Z3AlmostComplexComp}
\end{equation}
which encodes of the $\mathbb{Z}_3$-grading of the algebra $\mathfrak{g} = \mathfrak{h}\oplus \mathfrak{m}$. The fact that the eigenspaces of $\mathbb{J}$ do not close by construction hints towards the fact that this almost complex structure is not integrable. E.g. the torsion $T$ of the connection of associated almost Hermitan manifold $(M,g,\mathbb{J})$ is given by exactly that data, characterising the $\mathbb{Z}_3$-structure \cite{Butruille2005}: $T(m_1,m_2) \sim [m_1,m_2] \vert_\mathfrak{m}$.

Besides classifying the compact, six-dimensional nearly K\"ahler geometries, it has been shown in \cite{Butruille2005} as well that the almost complex structures on these examples, $S^3 \times S^3 = \frac{\text{SU}(2)\times\text{SU}(2)\times\text{SU}(2)}{\text{SU}(2)}$, $S^6 = \frac{G_2}{\text{SU}(3)}$, $\mathbb{CP}^3 = \frac{\text{Sp}(2)}{\text{SU}(2)\text{U}(1)}$ and the flag manifold $\frac{\text{SU}(3)}{\text{U}(1)\times\text{U}(1)}$, come from the $\mathbb{Z}_3$-symmetry.

Another possibility is, to have an \textbf{almost para-complex} resp. \textbf{nearly para-K\"ahler} structure on $M$. The difference to the almost complex case is mainly that here the eigenspaces $\mathfrak{g}^{(1)}$ and $\mathfrak{g}^{(2)}$ are subspaces to $\mathfrak{g}$ but the $\mathbb{Z}_3$-automorphism is only defined on $\mathfrak{g}^\mathbb{C}$. All formulas for that can be obtained from the ones above by $\mathbb{J} = i \mathbb{K}$ with an para-complex structure $\mathbb{K}^2 = 1$. Relations like \eqref{eq:Z3AlmostComplexComp} hold as well. An example for a para-complex $\mathbb{Z}_N$-coset and the integrable $\sigma$-model therein has been given recently in \cite{Delduc:2019lpe}.

\paragraph{The integrable $\sigma$-model in terms of geometric data.} In terms of an operator $\mathcal{O}$ with $\mathcal{O}^2 = \mathbb{1}$, which come either from a para-complex structure $\mathbb{K}$ or an almost complex structure $\mathbb{J}$ on $\mathfrak{m}$, with $\mathcal{O} = \mathbb{K}$ resp. $i\mathbb{J}$. In any case, $\mathcal{O} j^{(1)} = - j^{(1)}$ and $\mathcal{O} j^{(2)} = j^{(2)}$. Then, one can rewrite the $\mathbb{Z}_3$-coset action for a $\mathfrak{g}$-valued current $j = g^{-1} \mathrm{d} g$ from \eqref{eq:Lagrangian} as
\begin{align}
L &= \frac{1}{2} \text{Tr}\left( (s+b) j^{(1)}_+ j^{(2)}_- + (s-b) j^{(2)}_+ j^{(1)}_-  \right) = \frac{1}{2}\text{Tr}\left( m_+ (s + b \mathcal{O}) m_- \right) \nonumber\\
&\sim (s g + b \omega)(m_+,m_-)
\end{align}
for $m = j\vert_{\mathfrak{m}}$ and $h = j^{(0)} = j\vert_\mathfrak{h}$. The equations of motion $\mathcal{E} = 0$ and the Maurer-Cartan identity $\mathcal{M} = 0$ are computed as:
\begin{align}
\mathcal{E} &= \partial_+ m_- + \partial_- m_+ + [ h_+,m_-] - [m_+,h_-] - 3b\mathcal{O} [m_+ , m_-]\vert_{\mathfrak{m}} = 0 \nonumber \\
\mathcal{M}\vert_{\mathfrak{h}} &= \partial_+ h_- - \partial_- h_+ + [ h_+,h_-] + [m_+ , m_-]\vert_{\mathfrak{h}} = 0 \\
\mathcal{M}\vert_{\mathfrak{m}} &= \partial_+ m_- - \partial_- m_+ + [ h_+,m_-] + [m_+ , h_-] + [m_+ , m_-]\vert_{\mathfrak{m}} = 0, \nonumber
\end{align}
using the conditions \eqref{eq:Z3AlmostComplexComp}. Analogously to the previous section, the integrable choices for $s$ and $b$ is $b = \pm s/3$ and the Lax pair can be given only in terms of the decomposition $j = h + m$ and the structure $\mathcal{O}$:
\begin{align*}
L_+(\lambda) &= h_+ + \frac{1}{2}(\lambda^2 + \lambda)m_+ +  \frac{1}{2}( \lambda^2 - \lambda) \mathcal{O}m_+, \\
 L_-(\lambda) &= h_- + \frac{1}{2}(\frac{1}{\lambda^2} + \frac{1}{\lambda})m_- + \frac{1}{2}( \frac{1}{\lambda} - \frac{1}{\lambda^2}) \mathcal{O}m_-.
\end{align*}
The problem with the case of an almost complex structure is that the WZ-term $j^{(1)}\wedge j^{(2)}$ is imaginary due to $j^{(1)} = \overline{j^{(2)}}$. Hence, for the integrable $\sigma$-model, where the constant in front of that term is fixed, the Lagrangian corresponds to the pullback of some kind of hermitian metric $g + i\omega$ to the world-sheet. Such problems with reality were already pointed out in \cite{Young:2005jv}. For the para-complex case this problem does not occur and it is possible to write down a real action for integrable $\mathbb{Z}_3$-/$\mathbb{Z}_N$-models. 

Of course, one might be able to define other kind of integrable models on these nearly K\"ahler geometries. I.e. all of the earlier mentioned compact six-dimensional geometries admitting a $\mathbb{Z}_3$-grading corresponding to an almost complex structure -- $S^3 \times S^3$, $S^6$, $\mathbb{CP}^3$ and the flag manifold $\mathbb{F}^3$ -- are also $\mathbb{Z}_2$-symmetric homogeneous spaces. Hence, we can define $\mathbb{Z}_2$-cosets $\sigma$-models on those. Also, instead basing the construction on the $\mathbb{Z}_3$-grading (hence an almost complex structure), some of these also admit complex structures based on which interesting integrable models where defined in \cite{Bykov:2014efa,Bykov:2016rdv,Bykov:2014efa,Bykov:2018vbg,Bykov:2019jbz}. 

\paragraph{Example.} Let us study the case of the coset $G\times G \sim G\times G\times G/G_{diag}$, with a Lie (super)group $G$, to which also the case of $S^3 \times S^3$ belongs for $G = $ SU$(2)$. The $\mathbb{Z}_3$-outer automorphism $\sigma$ is simply the permutation of the three copies of $G$. One possible parameterisation of $G\times G \times G$ is $(g_1 h, g_2 h , h)$ for $g_i,h \in G$, such that $h$ parameterises $G_{diag}$ and can be subsequently chosen to be 1. 
\begin{align}
M = G\times G \hookrightarrow G \times G \times G,\ (g_1,g_2) \rightarrow (g_1,g_2,1)
\end{align}
This corresponds to the embedding $ \mathfrak{g} \times \mathfrak{g} = \mathfrak{m} \hookrightarrow  \mathfrak{g} \times \mathfrak{g}\times \mathfrak{g}$:
\begin{equation}
j = (j_1,j_2) = (g_1^{-1} \mathrm{d}g_1 , g_2^{-1} \mathrm{d}g_2) \rightarrow  (j_1,j_2,-j_1-j_2)
\end{equation}
on the level of the algebra. The currents $j_1 , j_2 \in \mathfrak{g}$, which are Maurer-Cartan forms on the two copies of $G$, should not be confused with the projections to the $\mathbb{Z}_3$-components $j^{(1)}$, $j^{(2)}$. A bilinear form on $\mathfrak{m}$ is inherited from the canonical Ad-invariant bilinear form on $\mathfrak{g} \times \mathfrak{g} \times \mathfrak{g}$ by that embedding, denoted by 'Tr',
\begin{equation} 
\left\langle (m_1,m_2),(n_1,n_2) \right\rangle = \text{Tr}\left(m_1 n_1 + m_2 n_2 + (m_1 + m_2)(n_1 + n_2) \right) .
\end{equation}
With that, the metric part of the Lagrangian becomes:
\begin{align}
L = \frac{1}{2} \left\langle j_+,j_- \right\rangle = \frac{1}{2} \left\langle (j_{1+},j_{2+}),(j_{1-},j_{2-}) \right\rangle \quad \Rightarrow \quad S \sim \int \text{Tr} \left( j_1 \wedge \star j_1 + j_2 \wedge \star j_2 + j_1 \wedge \star j_2 \right) . \nonumber
\end{align}
In particular this does \textit{not} correspond to the \textit{bi-invariant} metric on $G \times G$ anymore, because of the coupling term. But, as expected, it is invariant under the global action an element $(G_1,G_2,G_3) \in G \times G \times G$:
\begin{equation}
(G_1 , G_2 , G_3).(g_1,g_2) = (G_1 g_1 G_3, G_2 g_2 G_3), \quad (G_1 , G_2 , G_3).(j_1,j_2) = (G_3^{-1} j_1 G_3, G_3^{-1} j_2 G_3 )
\end{equation}
by the Ad$(G)$-invariance of the trace. The local symmetry $g_i \rightarrow g_i h(x)$ was fixed already by choosing the representatives on the currents $j \in \mathfrak{m}$ as above. This construction generalises to the construction of metrics on $G^N/G$-cosets with an $\mathbb{Z}_N$-automorphism \cite{Ledger1968}. 

From \eqref{eq:Z3ComplexStructureRel} one can derive a choice of almost complex structure $\mathbb{J}$ on $\mathfrak{m}$:
\begin{align*}
\mathbb{J} = \frac{1}{\sqrt{3}} \left( \begin{array}{cc} 1 & 2 \\ -2 & -1  \end{array} \right) , \qquad \mathbb{J}^2 = -1.
\end{align*}
Coupling the $\sigma$-model to corresponding 2-form of the compatible triple, $\omega = \left\langle \cdot , \mathbb{J} (\cdot) \right\rangle$:
\begin{equation}
L = \frac{1}{2} \left\langle j_+ , (\mathbb{1} + c\mathbb{J}) (j_-)\right\rangle \quad \Rightarrow \quad S \sim \int \text{Tr}\left( j_1 \wedge \star j_1 + j_2 \wedge \star j_2 + j_1 \wedge \star j_2 + \sqrt{3}cj_1 \wedge j_2 \right).
\end{equation}
This is a potentially interesting model of two interacting $\mathfrak{g}$-valued currents. But, as mentioned in the previous paragraph, the integrable choice of parameter of this model is imaginary $c = \pm \frac{i}{3}$. As a consequence, such a WZ-term gives an imaginary contribution to the action.

\subsubsection{$\mathbb{Z}_{N}$-supercosets as supergravity backgrounds}
From the point of view of string theory the question is, of course, whether $\mathbb{Z}_{2M}$-(super)-cosets could (for $M > 2$) give rise some new, probably quite exotic \footnote{due to the presence of $H$-flux}, supergravity backgrounds, potentially with AdS-factors for application in AdS/CFT. Compactifications based on nearly K\"ahler geometry have been discussed for example in \cite{LopesCardoso:2002vpf}.

Supercosets, where the isometry algebra is a Lie superalgebra with zero Killing form, and the parameters $b_i$ are chosen such that the $\sigma$-model is integrable (for $s_i = 1$), are supergravity backgrounds -- in the sense of them being conformally invariant at 1-loop. \cite{Kagan:2005wt}. This result is generalising the analysis for supergroup principal chiral models and $\mathbb{Z}_2$- and $\mathbb{Z}_4$-coset \cite{Bershadsky:1999hk,Berkovits:1999zq}.

The classical Lie superalgebras with zero Killing form are \cite{Kac:1978}, the basic Lie superalgebras $A(n,n)=\text{PSL}(n|n)$, $D(n+1|n)=\text{OSp}(2(n+1)|2n)$, the exceptional one $D(2|1;\alpha)$, and the strange Lie superalgebras $P(n)$, $Q(n)$ (both subalgebras of $\text{PSL}(n|n)$) and their real forms. For a review of their construction see \cite{Frappat:1996pb}. In principle, the possible outer $\mathbb{Z}_N$-automorphisms of these have been studied in \cite{Kac1969Automorphisms,Serganova1985} and should be scanned for potentially interesting cases that are applicable for supergravity backgrounds.

In addition to supersymmetric versions of the examples in the previous section, which were motivated by geometry, there other more group-theoretical types of gradings. Such are known in the construction of the strange Lie superalgebra $P(n)$ \cite{Kac:1978}, in the construction of harmonic superspace \cite{Galperin:2001seg}, or for several superconformal algebras several $\mathbb{Z}$-gradings appear, see for example \cite{Kac2004quantum}. In these examples the grading is basically a consequence of the Cartan decomposition and the grading comes from an \textit{inner automorphism} (the adjoint action with an Cartan element h) $\text{ad}(h)$. Take $\mathfrak{sl}(2)$ as an example with generators $(h,e_\pm)$ with $[h,e_\pm]=\pm e_\pm$, $[e_+,e_-]=h$. With identifications $\mathfrak{g}^{(0)} = \text{span}(h)$ and $\mathfrak{g}^{(1,2)} = \text{span}(e_\pm)$ there is a natural, but slightly trivial, $\mathbb{Z}_3$-grading due to $[\mathfrak{g}^{(i)},\mathfrak{g}^{(i)}] = 0$. The above mentioned grading on Lie superalgebras are generalisations of this structure.

\section{New integrable $\mathbb{Z}_N$-coset $\sigma$-models of hybrid type}
\label{chap:New}
So far, the literature on $\mathbb{Z}_N$-cosets has mainly centered around the pure spinor-type (super)string $\sigma$-models, reviewed in the previous section \cite{Young:2005jv,Kagan:2005wt,Kluson:2006ij,Bykov:2014efa,Bykov2016cyclic,Delduc:2019lpe}. But, Green-Schwarz superstring $\sigma$-models are important as generalisation of the Polyakov action on superspace, e.g. on AdS$_5  \times$S$^5$. Furthermore, as has been shown in \cite{Ke:2008zz} for $\mathbb{Z}_{4M}$-cosets in \cite{Hoare:2021dix} for general $\mathbb{Z}_N$-soets, certain Green-Schwarz like $\mathbb{Z}_N$-cosets $\sigma$-models exist for $N>4$, too.

The task is to explore this space of theories and to find constants $s_i$ (some of which are $0$) and $b_i$ such that the model \eqref{eq:Lagrangian} possesses a Lax representation. These are referred to as hybrid $\mathbb{Z}_N$-coset models in the following. The ansatz\footnote{chosen such that it automatically implies the Maurer-Cartan identity} for the Lax connection
\begin{equation}
L_+(\lambda) = j^{(0)}_+ + \sum_{i=1}^{N-1} l_i(\lambda) j^{(i)}_+, \qquad L_- = j^{(0)}_- + \sum_{i=1}^{N-1} \frac{1}{l_{N-i}(\lambda)} j^{(i)}_- \label{eq:LaxAnsatz}
\end{equation}
leads to the following equations for the constants $l_i$ when imposing both the flatness on $L$ and the equations of motion for the current $j = \mathfrak{g}^{-1} \mathrm{d} g$ \eqref{eq:ZNEOMDecomp}:
\begin{align}
i \text{ s.t. } s_i &= 0: \qquad l_{N-i} = \frac{1}{l_i}, \quad F_i(\lambda) \mathcal{D}_{ij} = \left\lbrace \begin{array}{cr}
l_j / l_{N+j-i} - l_i & i>j \\
l_j / l_{j-i} - l_i  & i<j \end{array}\right. \\
i \text{ s.t. } s_i &\neq 0: \qquad \frac{1}{2} \left(\frac{1}{l_{N-i}} - l_i \right) \frac{\mathcal{D}_{ij}}{s_i} = \left\lbrace\begin{array}{cr}
l_j / l_{N+j-i} - \frac{1}{2} \left( \frac{1}{l_{N-i}} + l_i \right) & i>j \\
l_j / l_{j-i} - \frac{1}{2} \left( \frac{1}{l_{N-i}} + l_i \right)  & i<j \end{array}\right.
\end{align}
with $\mathcal{D}_{ij}$ from \eqref{eq:ZNEOMDecompConst} given in terms of the constants $s_i$ and $b_i$. When this system is underdetermined the above ansatz is a spectral parameter dependent Lax connection. Because the 'equations of motion' for the $\mathfrak{g}^{(i)}$-component is a constraint when $s_i = 0$ the Lax condition has to give a constraint, as well. There can be an arbitrary spectral parameter dependent function of proportionality between these two constraints. This denoted above $F_i(\lambda)$. The freedom to choose those functions and the necessity for case separation makes the search for integrable choices of $s_i$ and $b_i$ difficult to handle. For that reason, only models for $N \leq 6$ are scanned in generality in the next section \ref{chap:Overview}, as $\mathbb{Z}_6$-cosets would potentially be interesting as supersymmetric version of the previously discussed $\mathbb{Z}_3$-cosets. A big class of models for arbitrary $N$, for which the constants $f_i$ can be given in a closed form, is presented in section \ref{chap:HybridClass}.

\subsection{Overview for $N \leq 6$} \label{chap:Overview}
In table \ref{tab:ListIntModelsToN=6} the integrable hybrid $\mathbb{Z}_N$-coset $\sigma$-models have been collected. The corresponding Lax connection is given there by specifying the constants $l_i$ in the ansatz \eqref{eq:LaxConnectionDefZ4DMV} as functions of the spectral parameter $\lambda$.
\begin{table}
$$\begin{array}{c|cc|c}
\text{grading} & \{s_i\} & \{b_i\} & \{l_i(\lambda) \} \\ \hline
\multirow{2}{*}{$\mathbb{Z}_3$} & (0,0) & (\frac{2}{3} , - \frac{2}{3}) & (\lambda , \frac{1}{\lambda}) \\
& (1,1) & ( - \frac{1}{3} , \frac{1}{3}) & (\lambda , \lambda^2) \\ \hline
\multirow{4}{*}{$\mathbb{Z}_4$} & (0,0,0) & \multirow{2}{*}{$(-b,0,b)$} & \multirow{2}{*}{no Lax pair found} \\  
& (1,0,1) &  & \\
& (0,1,0) & ( - \frac{1}{2},0, \frac{1}{2}) & ( \lambda , \lambda^2 , \frac{1}{\lambda} ) \\ 
& (1,1,1) & ( - \frac{1}{2},0, \frac{1}{2}) & ( \lambda , \lambda^2 , \lambda^3 ) \\ \hline
\multirow{4}{*}{$\mathbb{Z}_5$} & (0,0,0,0) & (-\frac{2}{5},-\frac{4}{5},\frac{2}{5},\frac{4}{5}) & (\lambda , \lambda^2 , \frac{1}{\lambda^2} , \frac{1}{\lambda}) \\ 
& (0,1,1,0) & ( - \frac{2}{5},\frac{1}{5},- \frac{1}{5},\frac{2}{5}) & ( \lambda , \lambda^2 , \lambda^3 \frac{1}{\lambda} ) \\ 
& (1,0,0,1) & ( - \frac{1}{5},-\frac{2}{5},\frac{2}{5},\frac{1}{5}) & (\lambda^3 , \lambda , \frac{1}{\lambda} , \lambda^2 )\\
& (1,1,1,1) & ( - \frac{3}{5}, - \frac{1}{5}, \frac{1}{5} , \frac{3}{5}) & ( \lambda , \lambda^2 , \lambda^3 , \lambda^4 ) \\ \hline
\multirow{8}{*}{$\mathbb{Z}_6$} & (0,0,0,0,0) & \multirow{4}{*}{$(-b_1,-b_2,0,b_2,b_1)$} & \multirow{4}{*}{no Lax pair found} \\ 
& (0,1,0,1,0) & & \\ 
& (1,0,1,0,1) & & \\
& (1,0,0,0,1) & & \\
& (0,0,1,0,0) & ( - \frac{1}{3}, - \frac{2}{3},0, \frac{2}{3},\frac{1}{3}) & ( \lambda , \lambda^2 , \lambda^3, \frac{1}{\lambda^2} , \frac{1}{\lambda} ) \\ 
& (0,1,1,1,0) & (-\frac{1}{3},\frac{1}{3},0,-\frac{1}{3},\frac{1}{3}) & (\lambda , \lambda^2 , \lambda^3 , \lambda^4 , \frac{1}{\lambda}) \\ 
& (1,1,0,1,1) & (-\frac{1}{3},\frac{1}{3},0,-\frac{1}{3},\frac{1}{3}) & (\lambda^2 , \lambda , 1, \lambda^2 , \lambda )\\
& (1,1,1,1,1) & ( - \frac{2}{3}, - \frac{1}{3}, 0 , \frac{1}{3} , \frac{2}{3}) & ( \lambda , \lambda^2 , \lambda^3 , \lambda^4 , \lambda^5 )
\end{array}$$
\caption{A, not necessarily complete, list for the data $\{s_i,b_i\}$ of $\mathbb{Z}_N$-coset $\sigma$-models for $N \leq 6$ and, if integrable and found by the author, the coefficients $\{l_i\}$ in the ansatz \eqref{eq:LaxAnsatz} of the corresponding Lax connection}
\label{tab:ListIntModelsToN=6}
\end{table}

All the integrable choices for $s_i$ and $b_i$, found there, have in common that all the non-zero $s_i$ are equal. As total factors in front of the action are neglected here, we choose $s_i = s_{N-i} = 1$ for the non-vanishing $s_i$. This equality might seem mysterious but it makes sense geometrically. Let $\tilde{\mathfrak{m}}$ denote $\sum_{i \ \text{with} \ s_i\neq 0} \mathfrak{g}^{(i)}$ then the kinetic term, coupling to some kind of metric, of the hybrid $\mathbb{Z}_N$-coset $\sigma$-model simply becomes
\begin{equation}
\mathcal{L} \sim \text{Tr} \left(j_+ \mathcal{P}_{\tilde{\mathfrak{m}}} (j_-) \right),
\end{equation}
which corresponds to the reduced metric to some kind of non-homogeneous quotient of $G$, as $(\mathfrak{g}-\tilde{\mathfrak{m}})$ does not correspond a subalgebra of $\mathfrak{g}$ in general. The maybe expected Green-Schwarz models for $s_i = \left\lbrace \begin{array}{cc} 1 & i \text{ even} \\ 0 & i \text{ odd} \end{array}\right.$ do not seem to exist for arbitrary $N$.

Furthermore, commenting on the found integrable choices for $b_i$, there might be more possibilities. For example, to all models we can find a second integrable model with $b_i \rightarrow - b_{i}$ and the corresponding Lax connection when exchanging $\mathfrak{g}^{(i)}$ and $\mathfrak{g}^{(N-i)}$.

\subsection{A class of hybrid $\mathbb{Z}_N$-coset $\sigma$-models}
\label{chap:HybridClass}
It seems illusory to find a closed form for all integrable choices of $b_i$ for arbitrary $N$, but one class of hybrid $\mathbb{Z}_N$-coset can be found, extrapolating
from table \ref{tab:ListIntModelsToN=6}. Let $N = 2M+1$ resp. $2M+2$, then a class of hybrid $\mathbb{Z_N}$-coset $\sigma$-models is characterised for each $N$ by a number $K = 1, ... , M$. The $\sigma$-model is defined by the choice of constants:
\begin{align*}
s_i &= 0 = s_{N-i} = 0 , \qquad b_{i} = - b_{N-i} = - \frac{2i}{N} \qquad \text{for} \quad i=1,...,K \\
s_i &= 1. \qquad b_i = 1 - \frac{2i}{N} \qquad \text{for} \quad i=K+1,...,N-K-1.
\end{align*}
So, the components $j^{(i)}$ and $j^{(N-i)}$ do not have kinetic terms in the action for $i=1,...,K$. A choice of Lax connection that is equivalent to the Maurer-Cartan identity and equations of motion is:
\begin{align}
L_+(\lambda) &= j_+^{(0)} + \sum_{i=1}^{N-K-1} \lambda^i j_+^{(i)} + \sum_{i=N-K}^{N-1} \frac{1}{\lambda^{N-i}} k_+^{(i)}  \label{eq:LaxZNhybridClass}\\
L_-(\lambda) &= j_-^{(0)} + \sum_{i=1}^{K1} \lambda^i j_-^{(i)} + \sum_{i=K+1}^{N-1} \frac{1}{\lambda^{N-i}} k_+^{(i)}  \nonumber
\end{align}
For $K = 0$ the action and original Lax connection of the pure spinor-type models \cite{Young:2005jv} is reproduced, for $N=4L$ and $K = M$ the Green-Schwarz action of \cite{Ke:2008zz}. The $K=M$ case for arbitrary $N$ has been constructed in \cite{Hoare:2021dix} already. As discussed there, it is noteworthy that for odd $N$ (as can also be seen for the $\mathbb{Z}_3$- and $\mathbb{Z}_5$-cases in table \ref{tab:ListIntModelsToN=6}), the $K=M$-case of the above class of models doesn't have any kinetic terms. Its dynamic is only governed by constraints. By means of the Lax connection \eqref{eq:LaxZNhybridClass} we have a Lax representation of this constraint structure. For the $\mathbb{Z}_3$-cosets based on an almost complex structure from section \ref{chap:Applications}, the action of that model could avoid the issues with reality of the action.

\section{Yang-Baxter deformed $\mathbb{Z}_N$-coset $\sigma$-models} \label{chap:YangBaxter}
\subsection{The action}
As the Yang-Baxter deformation of the $\mathbb{Z}_N$-model, we define
\begin{equation}
L = \frac{1}{2} \mathrm{Tr} \left( j_+ \mathcal{P}_- \frac{1}{1 - \eta R_g \circ \mathcal{P}_-} j_- \right) \label{eq:ActionZNDef}
\end{equation}
generalising the one of the principal chiral model, the $\mathbb{Z}_2$- and the $\mathbb{Z}_4$-(super)cosets. The deformation is characterised by a skewsymmetric\footnote{meaning $\mathrm{Tr}\left(m \ R(n) \right) = - \mathrm{Tr}\left(R(m) \ n \right)$. This is also true in case of a Lie superalgebra if, as is typically the case, the $R$-operator is chosen to be even.} operator $R: \ \mathfrak{g} \rightarrow \mathfrak{g}$ via the combination $R_g = \mathrm{Ad}_g^{-1} \circ R \circ \mathrm{Ad}_g$. $\eta$ is the deformation parameter, and the current $j$ and the projectors $\mathcal{P}_\pm$ are defined as in section \ref{chap:Review}.

The dynamics can be conveniently phrased in terms of the deformed currents
\begin{equation}
k_\pm = \frac{1}{1 \pm \eta R_g \circ \mathcal{P}_\pm} j_\pm . \label{eq:CurrentDef}
\end{equation}
The equations of motion are
\begin{equation}
\mathcal{E} = \partial_+ \left(\mathcal{P}_- k_- \right) + \partial_- \left(\mathcal{P}_+ k_+ \right) + [k_+ , \mathcal{P}_- k_-] + [k_-, \mathcal{P}_+ k_+] = 0 \label{eq:ZNDefEOM}
\end{equation}
and, in terms of the current $k$, the Maurer-Cartan equation for the current $j$ becomes
\begin{equation}
\mathcal{M} = \partial_+ k_- - \partial_- k_+ + [k_+ , k_-] + \eta^2 c^2 [\mathcal{P}_+ k_+ , \mathcal{P}_- k_-] - \eta R_g ( \mathcal{E} ) = 0, \label{eq:ZNDefMC}
\end{equation}
if the $R$-operator (and hence also the operator $R_g$) fulfils the modified classical Yang-Baxter equation:
\begin{equation}
[R(m) , R(n) ] - R\left([R(m),n] - [m , R(n)]\right) = c^2 [m,n]. \label{eq:MCYBE}
\end{equation}
All of these calculations work exactly analogously to \cite{Delduc:2013qra} for any choice of grading and any choice of projectors $\mathcal{P}_\pm$ (hence for any choice of constants $s_i$ and $b_i$). For two choices of $s_i$ and $b_i$, the Yang-Baxter deformation was already introduced in \cite{Hoare:2021dix}.

\paragraph{Homogeneous Yang-Baxter deformations.} We observe that, in case $R$ is a solution of the classical Yang-Baxter equation (\eqref{eq:MCYBE} for $c=0$), the equations of motion (hence also a potential Lax connection) will the same as in the undeformed case (as then $c \eta = 0$). In order to show that both models are really (classically) equivalent one has to check that the transformation $j \rightarrow k$ is a canonical transformation.

From the form of the action, resp. from the kind of redefinition \eqref{eq:CurrentDef} connecting deformed and undeformed model, we see that the deformation takes the form of a $\beta$-shift, generalising the deformation of the principal chiral model and the $\mathbb{Z}_2$- and $\mathbb{Z}_4$-case. In case $c = 0$, it is expected that this transformation is connected to abelian/non-abelian $T$-duality \cite{Matsumoto:2014nra,Osten:2016dvf,Hoare:2016wsk,Borsato:2016pas,Borsato:2017qsx,Fernandez-Melgarejo:2017oyu,Sakamoto:2017cpu,Hoare:2017ukq,Lust:2018jsx,Hoare:2018ngg}. The homogeneous Yang-Baxter deformation for the Green-Schwarz $\mathbb{Z}_{4M}$-(super)coset $\sigma$-model had already been constructed \cite{Ke:2017wis}. 

In the following this article will only be concerned with inhomogeneous Yang-Baxter (often called $\eta$-)deformations.

\subsection{$N=3$}
Let us the exemplify the construction of the Lax connection for the case of the $\mathbb{Z}_3$-coset. Decomposed into the components of the grading the Maurer-Cartan equation \eqref{eq:ZNDefEOM} and the equations of motion \eqref{eq:ZNDefMC} become
\begin{align}
\mathcal{M}^{(0)} &= \partial_+ k_-^{(0)} - \partial_- k_+^{(0)} + [k_+^{(0)} , k_-^{(0)}] + \mathcal{A}_1 [k_+^{(1)} , k_-^{(2)}] + \mathcal{A}_2 [k_+^{(2)} , k_-^{(1)}] = 0  \nonumber \\
\mathcal{M}^{(1)} &= \partial_+ k_-^{(1)} - \partial_- k_+^{(1)} + [k_+^{(0)} , k_-^{(1)}] + [k_+^{(1)} , k_-^{(0)}] + \mathcal{C} [k_+^{(2)} , k_-^{(2)}] = 0 \nonumber \\
\mathcal{M}^{(2)} &= \partial_+ k_-^{(2)} - \partial_- k_+^{(2)} + [k_+^{(0)} , k_-^{(2)}] + [k_+^{(2)} , k_-^{(0)}] + \mathcal{C} [k_+^{(1)} , k_-^{(1)}] = 0 \label{eq:Z3DefEOM}\\
(\mathcal{E}^{(1)} + b \mathcal{M}^{(1)})/s &= \partial_+ k_-^{(1)} + \partial_- k_+^{(1)} + [k_+^{(0)} , k_-^{(1)}] - [k_+^{(1)} , k_-^{(0)}] - \mathcal{D} [k_+^{(2)} , k_-^{(2)}] = 0 \nonumber \\
(\mathcal{E}^{(2)} - b \mathcal{M}^{(2)})/s &= \partial_+ k_-^{(2)} + \partial_- k_+^{(2)} + [k_+^{(0)} , k_-^{(2)}] - [k_+^{(2)} , k_-^{(0)}] + \mathcal{D} [k_+^{(1)} , k_-^{(1)}] = 0 \nonumber
\end{align}
with
\begin{align*}
\mathcal{A}_1 &= 1 + \epsilon^2 ( s - b )^2, \quad \mathcal{A}_2 = 1 + \epsilon^2 ( s + b )^2 , \\
\mathcal{C} &= 1 + \epsilon^2 ( s^2 - b^2 ) \quad \text{and} \quad \mathcal{D} = \frac{b}{s}\left(3 + \epsilon ( s^2 - b^2 ) \right) = \frac{b}{s}\left(2 - \mathcal{C} \right).
\end{align*}
where we defined $\epsilon = c\eta$ and
Let us take the following ansatz for the Lax connection:
\begin{equation}
L_\pm  = k_\pm^{(0)} + l_{\pm,1} k_\pm^{(1)} + l_{\pm,2} k_\pm^{(2)}. \nonumber
\end{equation}
One ends up with the following four equations for coefficients $l_{\pm,i}$
\begin{align*}
l_{+,1} \ l_{-,1} &= \frac{l_{-,2} + l_{+,2}}{2} \cdot \mathcal{C} +  \frac{l_{-,2} - l_{+,2}}{2} \cdot \mathcal{D}, \qquad l_{+,1} \ l_{-,2} = \mathcal{A}_1 \\
l_{+,2}\ l_{-,2} &= \frac{l_{-,1} + l_{+,1}}{2} \cdot \mathcal{C} -  \frac{l_{-,1} - l_{+,1}}{2} \cdot \mathcal{D}, \qquad l_{+,2} \ l_{-,1} = \mathcal{A}_2
\end{align*}
when imposing the Lax condition $\partial_+ L_- - \partial_- L_+ + [L_+ , L_-] = 0$. In order to make that system underdetermined, it is sufficient to impose the following two conditions on $\mathcal{A}_1$, $\mathcal{A}_2$, $\mathcal{C}$ and $\mathcal{D}$, which in turn allow to determine the deformation of the constants $s$ and $b$ in terms of $\epsilon$:
\begin{align}
\mathcal{C} = \mathcal{D} \qquad \text{and} \qquad \mathcal{A}_2 = \mathcal{C}^2.
\end{align}
Both equations are solved simultaneously by
\begin{equation}
\epsilon^2 = \frac{3b-s}{(s-b)^2 (s+b)} \label{eq:SolDefParZ3} 
\end{equation}
allowing to eliminate $\mathcal{C}$ and $\mathcal{D}$, and determine constants $s$ and $b$ in terms of $\epsilon$. There will be a more complicated relation between $s$ and $b$ than $b = \pm s/3$ in the undeformed model. Still, there is a remaining degree of freedom for the choice of $s$ and $b$, which is analogous to the freedom to rescale both $s$ and $b$ in the undeformed model. Potentially, this remaining degree of freedom will be fixed in a Hamiltonian analysis, when requiring that the associated charges are in involution. 

With the above conditions, we get that $l_{+,2} = \sqrt{\mathcal{A}_2} (l_{+,1}/\sqrt{\mathcal{A}_1})^2$. The Lax connection with spectral parameter $\lambda$ takes the form:
\begin{equation}
L_+(\lambda) = k_+^{(0)} + \lambda \sqrt{\mathcal{A}_1} k_+^{(1)} + \lambda^2 \sqrt{\mathcal{A}_2}  k_+^{(2)}, \qquad L_-(\lambda) = k_-^{(0)} + \frac{1}{\lambda^2} \sqrt{\mathcal{A}_2} k_-^{(1)} + \frac{1}{\lambda} \sqrt{\mathcal{A}_1} k_+^{(2)} \label{eq:LaxPairDefZ3Young}
\end{equation}
with the constants $s$, $b$ in $\mathcal{A}_i$ (partially) fixed by equation \eqref{eq:SolDefParZ3}. This form will motivate the ansatz for the deformed $\mathbb{Z}_N$-models.\footnote{Let us note that there is a second set of sufficient conditions:
\begin{align}
\mathcal{C} = - \mathcal{D} \qquad \text{and} \qquad \mathcal{A}_1 = \mathcal{C}^2. \nonumber
\end{align}
This leads to a second integrable model, analogous to the undeformed model. The corresponding identifications are still $b_i \leftrightarrow b_{N-i}=-b_i$ and $\mathfrak{g}^{(i)} \leftrightarrow \mathfrak{g}^{(N-i)}$.} 

\subsection{General $N$}
The Maurer-Cartan equation \eqref{eq:ZNDefMC} and the equations of motion \eqref{eq:ZNDefEOM} become
\begin{align}
\mathcal{M}^{(0)} &= \partial_+ k_-^{(0)} - \partial_- k_+^{(0)} + [k_+^{(0)} , k_-^{(0)}] + \sum_{j=1}^{N-1} \mathcal{A}_j [k_+^{(j)} , k_-^{(N-j)}] = 0  \nonumber \\
\mathcal{M}^{(i)} &= \partial_+ k_-^{(i)} - \partial_- k_+^{(i)} + [k_+^{(0)} , k_-^{(i)}] + [k_+^{(i)} , k_-^{(0)}] + \sum_{j\neq 0,i} \mathcal{C}_{ij} [k_+^{(j)} , k_-^{(i-j)}] = 0 \label{eq:ZNDefEOMDecomp}\\
(\mathcal{E}^{(i)} + b_i \mathcal{M}^{(i)}) &= s_i (\partial_+ k_-^{(i)} + \partial_- k_+^{(i)} + [k_+^{(0)} , k_-^{(i)}] - [k_+^{(i)} , k_-^{(0)}]) + \sum_{j\neq 0,i} \mathcal{D}_{ij} [k_+^{(j)} , k_-^{(i-j)}] = 0 \nonumber
\end{align}
when decomposed in the grading eigenspaces ($i = 1,...,N-1$). The constants $\mathcal{A}_j$, $\mathcal{C}_{ij}$ and $\mathcal{D}_{ij}$ are:
\begin{align}
\mathcal{A}_i &= 1 + \epsilon f_i^2, \quad \mathcal{C}_{ij} = 1 +  \epsilon f_j f_{j-i} , \quad \mathcal{D}_{ij} = f_{j-i} - f_{j} - b_i \mathcal{C}_{ij} \label{eq:ZNDefEOMDecompConstants}
\end{align}
where again $\epsilon = c \eta$ was introduced and the indices on $s$ and $b$ are to be understood as 'mod $N$', so e.g. $b_{-i} = b_{N-i} = - b_i$ for simplicity of notation.

\subsubsection{Deformed pure spinor-type $\mathbb{Z}_N$-coset $\sigma$-models}
Motivated by the 'pure spinor' $\mathbb{Z}_3$-case, we make the following ansatz for the Lax connection in case $s_i \neq 0$.
\begin{equation}
L_+(\lambda) = k_+^{(0)} + \sum_{i=1}^{N-1} \lambda^i \sqrt{\mathcal{A}_i} k_+^{(i)}, \qquad L_-(\lambda) = k_-^{(0)} + \sum_{i=1}^{N-1} \frac{1}{\lambda^{N-i}} \sqrt{\mathcal{A}_{N-i}} k_-^{(i)} \label{eq:LaxConnectionZNDefSpinor}
\end{equation}
Indeed the flatness condition to this Lax connection is equivalent to the equations of motion and Maurer-Cartan identity above \eqref{eq:ZNDefEOMDecomp}, if the following equations hold:
\begin{align}
\mathcal{C}_{ij}^2 = \left\lbrace \begin{array}{cr}
\frac{\mathcal{A}_j \mathcal{A}_{N+j-i}}{\mathcal{A}_{N-i}} & i>j \\
\frac{\mathcal{A}_j \mathcal{A}_{j-i}}{\mathcal{A}_{i}}  & i<j
\end{array} \right. \\
\mathcal{D}_{ij} = \left\lbrace \begin{array}{cr}
s_i \mathcal{C}_{ij} & i>j \\
- s_i \mathcal{C}_{ij}  & i<j
\end{array} \right.
\end{align}
These equations are solved when the constants $f_i = s_i - b_i$ solve the following equations
\begin{equation}
- \epsilon^2 =  \left\lbrace \begin{array}{cr}
\frac{f_{N+j-i} - f_{j} + f_{N-i}}{f_{N+j-i} f_j f_{N-i}} & i>j \\
\frac{f_{j-i} - f_{j} + f_{i}}{f_{j-i} f_j f_i}  & i<j
\end{array} \right. . \label{eq:SolDefParZNDef}
\end{equation}
A solution to this condition is
\begin{equation}
f_i = \frac{1}{\epsilon} \tan\left( \sigma(\epsilon) \epsilon \frac{2i}{N} \right) \qquad \overset{\epsilon \rightarrow 0}{\longrightarrow} \quad \sigma(0) \frac{2i}{N} \label{eq:SolFiZNDef}
\end{equation}
with $\epsilon = c\eta$ and some regular function $\sigma(\epsilon)$ of the deformation parameter. We observe that this gives the expected form for $f_i = s_i - b_i$ in the undeformed limit, $f_i(\epsilon = 0) = \sigma(0) \frac{2i}{N}$, with $\sigma(0)$ playing the role a total scale there. The freedom to choose this function is the same freedom that was observed in the $\mathbb{Z}_3$-model. This function would be expected to be fixed when working in the Hamiltonian formalism, analogously to the $\mathbb{Z}_4$-case \cite{Delduc:2014kha,Lacroix:2018njs}. In the scope of this paper, we will determine it below by comparing it to the known Green-Schwarz $\mathbb{Z}_4$-case \cite{Delduc:2013qra}.

Resultantly, \eqref{eq:LaxConnectionZNDefSpinor} is a Lax connection if $f_i$ is given by \eqref{eq:SolFiZNDef}. Let us give explicit expressions for $\mathcal{A}_i$ and $s_i$ in terms of the deformation parameter that will be useful in the following:
\begin{align}
\mathcal{A}_i &= \frac{1}{\cos^2\left(\sigma(\epsilon) \epsilon \frac{2i}{N}\right)}, \qquad s_i = \frac{1}{2 \epsilon} \sin\left( 2 \epsilon \sigma(\epsilon) \right) \sqrt{\mathcal{A}_i \mathcal{A}_{N-i}}. \label{eq:SolAiZNDef}
\end{align}

\subsubsection{Deformed hybrid $\mathbb{Z}_N$-coset $\sigma$-models with $s_i = 0$ for $i\leq K$}
As in section \ref{chap:New} one generalisation of the above in detail is discussed in detail. We use the same conditions as there. But, analogously to the Yang-Baxter-deformed 'pure spinor'-type $\mathbb{Z}_N$-model, the constants $s_i$, $b_i$ defining the models need to be deformed, too.
\begin{align}
f_{\bar{I}} &= - f_{N - \bar{I}} = \frac{1}{\epsilon} \tan \left( \sigma(\epsilon) \epsilon \frac{2\bar{I}}{N} \right) , \qquad f_{I} =  \frac{1}{\epsilon} \tan \left( \sigma(\epsilon) \epsilon \frac{2I}{N} \right)
\label{eq:DefZNConstants}
\end{align}
This corresponds to:
\begin{align}
\mathcal{A}_{\bar{I}} &= \mathcal{A}_{N - \bar{I}} = \frac{1}{\cos \left( \sigma(\epsilon) \epsilon \frac{2\bar{I}}{N} \right)} , \qquad \mathcal{A}_{I} =  \frac{1}{\cos \left( \sigma(\epsilon) \epsilon \frac{2\bar{I}}{N} \right)} \\
s_{\bar{I}} &= s_{N-\bar{I}} = 0, \qquad s_I = \frac{1}{2 \epsilon} \sin\left( 2 \epsilon \sigma(\epsilon) \right) \sqrt{\mathcal{A}_I \mathcal{A}_{N-I}} \label{eq:DefZNConstants2}
\end{align}
A Lax pair, combining the ideas of the deformed pure spinor-type $\mathbb{Z}_N$-model and the one of hybrid hybrid $\mathbb{Z}_N$-coset models, is this one
\begin{align}
L_+(\lambda) &= k_+^{(0)} + \sum_{i=1}^{N-K-1} \lambda^i \sqrt{\mathcal{A}_i} k_+^{(i)} + \sum_{i = N-K}^{N-1} \frac{\sqrt{\mathcal{A}_i}}{\lambda^{N-i}} k_+^{(i)} \nonumber \\
L_-(\lambda) &= k_-^{(0)} + \sum_{i=1}^{K} \lambda^i \sqrt{\mathcal{A}_{N-i}} k_-^{(i)} + \sum_{i = K+1}^{N-1} \frac{\sqrt{\mathcal{A}_{N-i}}}{\lambda^{N-i}} k_-^{(i)} \label{eq:LaxConnectionDefZNClassK} \\
&= k_-^{(0)} + \sum_{i=1}^{N-K-1} \frac{\sqrt{\mathcal{A}_i}}{\lambda^i} k_-^{(N-i)} + \sum_{i = N-K-1}^{N-1} \lambda^{N-i} \sqrt{\mathcal{A}_i} k_-^{(N-i)} . 
\end{align}
Inserting this into the flatness condition $\partial_+ L_- - \partial_- L_+ + [L_+ , L_-] = 0$ and using equation of motion and Maurer-Cartan equations \eqref{eq:ZNDefEOMDecomp} results in the following conditions on the constants $\mathcal{A}_i$, $\mathcal{C}_{ij}$ and $\mathcal{D}_{ij}$ from \eqref{eq:ZNDefEOMDecompConstants} for the $\mathfrak{g}^{(i)}$-component of the flatness condition:
\begin{itemize}
\item \underline{$i\leq K$ or $i\geq N-K$}:
In two cases a non-trivial condition arises
\begin{align}
F_i(\lambda)\mathcal{D}_{ij} = - \left\lbrace \begin{array}{lc} \lambda^i \left( \sqrt{\mathcal{A}_j \mathcal{A}_{N+j-i}}/\lambda^N - \mathcal{C}_{ij} \sqrt{A}_i \right) & \quad i \leq K, \ j \geq N-K, \ j-i \leq N-K-1 \\
\lambda^i \left( \sqrt{\mathcal{A}_j \mathcal{A}_{j-i}} - \mathcal{C}_{ij} \sqrt{A}_i/\lambda^N \right) & \quad i,j \geq N-K, \ i-j \leq K
\end{array} \right.
\end{align}
In both cases the corresponding functions of proportionality are
\begin{equation}
F_i(\lambda) = \frac{2\epsilon}{\sin \left(2\epsilon \sigma(\epsilon)\right)} \frac{ \cos\left(\sigma(\epsilon) \epsilon  \frac{2(N-i)}{N} \right) \lambda^N - \cos\left(\sigma(\epsilon) \epsilon  \frac{2i}{N}\right)}{2 \lambda^{N-i}}.
\end{equation}
Again, as in the undeformed case, this insures that constraint structure arising from the flatness of the Lax connection is the same one coming from the equation of motion, as again the $\mathfrak{g}^{(i)}$-component of the equation of motion \eqref{eq:ZNDefEOMDecomp} for $s_i = 0$ is a constraint. These constraints are:
\begin{equation}
0 \approx \Phi^{i} = \left\lbrace \begin{array}{lc} \sum_{j=N-K}^{N-K-1+i} \mathcal{C}_{ij} \sqrt{\mathcal{A}_i} [k_+^{(j)},k_-^{(N+i-j)}]  & \quad i=\bar{I} \leq K \\
\sum_{j=i-K-1}^{N-K-1} \mathcal{C}_{ij} \sqrt{\mathcal{A}_i} [k_+^{(j)},k_-^{(i-j)}]  & \quad i=N - \bar{I} \geq N-K
\end{array} \right.
\end{equation}
The concrete form of the $F_i(\lambda)$ is, of course, not important here but it is important to show that it is exists and it is also notworthy that similar functions appeared in the Hamiltonian analysis of the $\mathbb{Z}_4$ case with constraints \cite{Magro:2008dv,Vicedo:2009sn}. All other contributions to the constraints vanish when the following conditions are fulfilled 
\begin{align*}
\mathcal{D}_{ij} = 0, \qquad \mathcal{C}^2_{ij}= \left\lbrace \begin{array}{cr}
\mathcal{A}_j \mathcal{A}_{N+j-i}/\mathcal{A}_i & i>j \\
\mathcal{A}_j \mathcal{A}_{j-i}/\mathcal{A}_i & i < j
\end{array} \right. ,
\end{align*}
keeping in mind, that here $\mathcal{A}_i = \mathcal{A}_{N-i}$.

\item \underline{$K+1 \leq i \leq N-K-1$}:
The two relevant sets of equation, that arise here when in comparing the flatness of the Lax connection to equations of motion and Maurer-Cartan equation \eqref{eq:ZNDefEOMDecomp} are:
\begin{align}
0 &= \left\lbrace \begin{array}{lc}
\mathcal{D}_{ij} + s_i \mathcal{C}_{ij} = f_{N+j-i} - f_j + f_i \mathcal{C}_{ij} & \quad 0< i -j \leq K, \ j \leq N-K-1 \\
\mathcal{D}_{ij} - s_i \mathcal{C}_{ij} = f_{N+j-i} - f_j - f_{N-i} \mathcal{C}_{ij} & \quad K+1\leq i -j \leq N-K-1, \ j \leq N-K-1 \\
\mathcal{D}_{ij} + s_i \mathcal{C}_{ij} = f_{j-i} - f_j + f_i \mathcal{C}_{ij} & \quad j \leq N- K -1,\  0 < j-i\leq N-K-1 \\
\mathcal{D}_{ij} - s_i \mathcal{C}_{ij} = f_{j-i} - f_j + f_{N-i} \mathcal{C}_{ij} & \quad j \geq N-K, \ 0 < j-i \leq N-K-1
\end{array} \right.
\end{align}
and
\begin{align}
\mathcal{C}_{ij}^2 = \left\lbrace \begin{array}{cl}
\mathcal{A}_j \mathcal{A}_{N+j-i} / \mathcal{A}_{N-i} & \quad 0< i -j \leq K, \ j \leq N-K-1 \\
\mathcal{A}_j \mathcal{A}_{N+j-i} / \mathcal{A}_{i} & \quad K+1\leq i -j \leq N-K-1, \ j \leq N-K-1 \\
\mathcal{A}_j \mathcal{A}_{j-i} / \mathcal{A}_{i} & \quad j \leq N- K -1,\  0 < j-i\leq N-K-1 \\
\mathcal{A}_j \mathcal{A}_{j-i} / \mathcal{A}_{N-i} & \quad j \geq N-K, \ 0 < j-i \leq N-K-1
\end{array} \right.
\end{align}

\end{itemize}
With choice of constants $f_i$ in \eqref{eq:DefZNConstants}, and resultantly for the $\mathcal{A}_j$ in, \eqref{eq:DefZNConstants2} all the above conditions are fulfilled. Hence, \eqref{eq:LaxConnectionDefZNClassK} is the Lax representation of the Yang-Baxter deformed partly Green-Schwarz type $\mathbb{Z}_N$-graded (super)coset $\sigma$-model

\paragraph{Fixing the function $\sigma(\epsilon)$.} The $N=4$, $K=1$-model is the Green-Schwarz type $\mathbb{Z}_4$-supercoset. Let us compare the constants to the Lax connection of Delduc, Magro and Vicedo \cite{Delduc:2013qra}:
\begin{equation}
L^{\text{(DMV)}}_\pm(\lambda) = k_\pm^{(0)} + \lambda \sqrt{1+\epsilon^2}  k_\pm^{(1)} + \lambda^{\pm2} \frac{1 + \epsilon^2}{1-\epsilon^2}  k_\pm^{(1)} + \frac{1}{\lambda} \sqrt{1+\epsilon^2} k_\pm^{(3)} \label{eq:LaxConnectionDefZ4DMV}
\end{equation}
and the projector $\mathcal{P}_- = - \mathcal{P}^{(1)} + \frac{2}{1-\epsilon^2} \mathcal{P}^{(2)} + \mathcal{P}^{(3)}$. The corresponding identifications are
\begin{align*}
\sqrt{\mathcal{A}_2} &= \frac{1}{\cos \left(\sigma(\epsilon) \epsilon \right)} \overset{!}{=} \frac{1+\epsilon^2}{1-\epsilon^2} \quad \Rightarrow \quad \sigma(\epsilon) = \frac{1}{\epsilon} \arccos\left(\frac{1-\epsilon^2}{1+\epsilon^2}\right)
\end{align*}
which implies $\sin(\epsilon \sigma(\epsilon)) = \frac{2\epsilon}{1+\epsilon^2}$, $\sin(2\epsilon \sigma(\epsilon)) = 4\epsilon \frac{1-\epsilon^2}{(1+\epsilon^2)^2}$ and, as wished,
\begin{align*}
\sqrt{\mathcal{A}_1} &= \sqrt{\mathcal{A}_3} = \frac{1}{\cos \left(\sigma(\epsilon) \epsilon/2 \right)} = \sqrt{1+\epsilon^2}, \\ 
s_2 &= \frac{1}{\epsilon} \tan(\epsilon\sigma(\epsilon)) = \frac{2}{1 - \epsilon^2}, \qquad b_1 = -\frac{1}{\epsilon} \tan(\epsilon\sigma(\epsilon)/2) = - 1,
\end{align*}
showing that indeed \eqref{eq:LaxConnectionDefZNClassK} is a generalisation of \eqref{eq:LaxConnectionDefZ4DMV} for arbitrary deformed $\mathbb{Z}_N$-(super)-coset $\sigma$-models, valid for each $K=1,...,M$. This choice of function $\sigma(\epsilon)$ should also reproduce the found actions for $K=0$ and $K=M$ from \cite{Hoare:2021dix}.

But let it be noted that it is not necessary to fix the function $\sigma(\epsilon)$ from the point of view of equivalence of the flatness of the Lax connection to the equations of motion and Maurer-Cartan equation. It might be necessary to fix it to ensure that the associated conserved charges are in involution. Also, in the Lax formulation in terms of the deformed currents (and typically also in the associated deformation of the Poisson structure) only the combination $\epsilon^2 = c^2 \eta^2$ appears. So, in principle, both non-split ($c=1$) and split solutions ($c = i$) to the modified classical Yang-Baxter deformation seem possible in defining a deformation and keeping things real. In case of $c=i$, one can set $\epsilon^2 \rightarrow - \eta^2$. $c$ only appears linearly in the arguments of the expressions for $\mathcal{A}_i$, $f_i$, etc., in every instance as argument of a trigonometric function. 

\section{Outlook}
Several new models were proposed in this article together with their Lax representations. In order to complete the proof of classical integrability of these models a Hamiltonian analysis should be performed. It is expected that one needs to invoke $r-s$-formalism and that both the undeformed and deformed $\mathbb{Z}_N$-models fall into the class of models with twist function \cite{Ke:2011zzb,Ke:2011zzc,Ke:2011zzd,Lacroix:2018njs,Delduc:2019lpe}. This would show the involution of the conserved charges associated to these Lax pairs. Moreover, in case of the hybrid models a thorough constraint analysis is needed to show Hamiltonian integrability, i.e. involution of the conserved charges and all the constraints, of this class of models. In the $\mathbb{Z}_4$-supercoset Green-Schwarz $\sigma$-model the first class part of the (fermionic) constraints corresponding to the non-existent 'kinetic' term for $j^{(1)}$ and $j^{(3)}$ is the generator for the (fermionic) $\kappa$-symmetry \cite{Vicedo:2009sn}, raising the question what happens when $s_i$ is a bosonic direction. Also, taking analogy from the  Green-Schwarz $\mathbb{Z}_4$-supercoset $\sigma$-model where the WZ-term allows for $\kappa$-symmetry, which in turn allows get rid of some fermionic degrees of freedom, one would assume that this $\sigma$-model has some kind of effective target space, in which some degrees of freedom associated to the components $j^{(i)}$ for which $s_i = 0$ do not contribute.

This might also clarify the relationship between different types of $\mathbb{Z}_N$-coset $\sigma$-model for the same $N$ and the same coset $M = G/H$ but different choice of constants $s_i$. In analogy to the pure spinor and Green-Schwarz superstring on $\mathbb{Z}_4$-cosets, which are different formulations of the same theory, one might expect some relation between such models. Another question is what the relationship between two $\sigma$-models on $M = G/H$ resp. $M=\tilde{G}/\tilde{H}$, where $M$ can be respresented by two different cosets with $\mathbb{Z}_N$ symmetry, but with different $N$. For some cases it has been shown in \cite{Bykov2016cyclic} that these models differ only by a topological term in the action and hence are classically equivalent. It would be interesting to investigate that further.

In the end the answer to the question, how useful these integrable models could be in the study of string theory, depends on whether one finds some homogeneous supergravity backgrounds that possess $\mathbb{Z}_N$-symmetry. These seem to be quite exotic but related geometries, like nearly K\"ahler geometries have been discussed in the context AdS-flux compactifications \cite{LopesCardoso:2002vpf} supergroup cosets leading to supergravity solutions, i.e. if the isometry group of the homogeneous space is a supergroup with vanishing Killing form \cite{Kagan:2005wt} might be possible. Analogously to \cite{Borsato:2016ose} it should be also check whether for the Yang-Baxter deformation some kind of unimodularity condition on the operator $R$ ensures is sufficient such that the deformed background is still a solution of supergravity.

\subsection*{Acknowledgements}
The author thanks Dmitri Bykov, Ben Hoare, Nat Levine and Maxime Trepanier for discussions and Arkady Tseytlin for comments on the draft.

\bibliographystyle{jhep}
\bibliography{References}

\end{document}